# Negative differential resistance and pulsed current induced multi-level resistivity switching in charge ordered and disordered manganites

A. Rebello and R. Mahendiran

Department of Physics, Faculty of Science, National University of Singapore,

2 Science Drive 3,   Singapore -117542

**Abstract**

We investigated pulsed and direct (*dc*) current-induced electroresistance in two insulators of different electronic and magnetic ground states: charge-orbital ordered antiferromagnetic $Nd_{0.5}Ca_{0.5}MnO_3$ (NCMO) and charge disordered ferromagnetic $LaNi_{0.5}Mn_{0.5}O_3$ (LNMO). A systematic study of *dc* and pulsed current–voltage (*I-V*) characteristics of varying pulse width or period along with simultaneous temperature measurements suggest that hysteresis and negative differential resistance (NDR) observed at high current densities during *dc* current sweep in these compounds vanish systematically with increasing period of the pulsed current, thus underlining the importance of Joule heating in the NDR phenomenon. In the low current regime with negligible Joule heating, bi-level and multi-level resistivity ($\rho$) switching induced by a sequence of pulses with controlled pulse width or period with a fixed amplitude are shown. It is found that $\rho$ increases abruptly ($\approx$ 37 % for NCMO at 100 K and $\approx$ 17 % for LNMO at 300 K) upon increase of pulse period from 50 ms to 100 ms for a fixed pulse width (25 ms) and current (*I* = 2 mA). Similar resistive switching effects of different magnitudes were also found with variations in the pulse width for a fixed period and our results suggest that these materials can be exploited for non-volatile memory based on resistivity. We discuss possible origins of the observed pulse width and pulse period controlled electroresistance effect.

Electrically addressable non-volatile memory devices are currently attracting enormous attention due to their potential applications in compact data storage.[1] In particular, a large change of electrical resistance on application of an external *dc* electric field, i.e. colossal electroresistance (CER), observed in charge ordered manganites such as $Pr_{0.7}Ca_{0.3}MnO_3$ triggered a surge of interest into potential applications of these classes of materials as non-volatile memory (NVM) elements in resistive random access memories (RRAM).[2] The CER effect has been seen in single crystals,[3] thin films[4] and polycrystalline materials[5] and it can occur within the bulk of the material or at the electrode-oxide interface. Various mechanism such as electric field- induced melting of charge ordering,[2] creation of metallic filamentary channels and their percolation in the insulating matrix,[4,6] excitation of charge density waves,[7] small-polaron-hopping,[8] and migration of oxygen ions,[9] etc., have been suggested as possible origins associated with the bulk of the sample but a clear consensus on the exact mechanism of the CER phenomenon is yet to emerge. Generally, CER associated with the bulk mechanisms starts to appear below a characteristic temperature, for example, charge ordering temperature in manganites in contrast to the interfacial CER which occurs even at room temperature in metal-oxide-metal sandwich structures due to the electrical field induced changes in the contact resistance.[1,10,11] Interestingly, polarity dependent resistivity switching, i.e., high resistance state ("OFF" state) for positive voltage pulses and low resistance ("ON" state) for negative voltage pulses in $Pr_{0.67}Ca_{0.33}MnO_3$ capacitor-like structures was also demonstrated recently.[1,9,12] The magnitude of electroresistance was found to depend on the polarity, amplitude and number of pulses. However, the influence of pulse period and pulse width on the CER was rarely investigated except in few mangaites. Odagawa et al.[13] shown that CER in $Pr_{0.7}Ca_{0.3}MnO_3$ thin film is independent on the duration (width) of pulses from an infinite (dc) down 150 ns. However, it was two probe measurements in metal-oxide-metal sandwich structure in which charge-trapping with the Schottky barrier region at the metal-oxide interface plays important role.

One of the central issues related to the CER phenomenon is the origin of the negative differential resistance (NDR) which appears above a threshold current or voltage during *dc* current or voltage sweep. Its origin has been attributed to opening of metallic filaments,[4] depinning of charge ordering[13,15] or polarons[7] but the role of self- Joule heating was not seriously considered. There are diverging opinions on the role and influence of Joule heating [6,14,15,16,17,18] and therefore this issue has to be addressed before attributing the observed CER to any new electronic mechanisms. Moreover, majority of the bulk related CER work reported so far are on the charge-orbital ordered compounds or phase segregated compounds showing semiconductor-metal transition as a function of temperature. Detailed investigations in other compounds may shed light on the roles of charge-orbital ordering and microscopic phase separation on the CER effect.

The two main objectives of this report are to investigate the origin of the negative differential resistance and to explore the possibility of pulsed current induced resistivity switching in two half-doped manganites with different magnetic and electronic ground states: antiferromagnetic charge-orbital ordered insulator $Nd_{0.5}Ca_{0.5}MnO_3$ (NCMO) in which $e_g$-electrons of $Mn^{3+}$ :$t_{2g}^3 e_g^1$ and holes of $Mn^{4+}$ :$t_{2g}^3 e_g^0$ and $e_g$-orbitals ordered below 240 K[19], and charge-orbital disordered insulator $LaNi_{0.5}Mn_{0.5}O_3$ (LNMO) which is ferromagnetic below 275 K due to ferromagnetic superexchange interaction between B-site disordered $Ni^{2+}$ and $Mn^{4+}$ ions.[20] The latter compound does not suffer from the problem of microscopic phase separation which is commonly encountered in other manganites. The LNMO compound has recently aroused interest due to the reported magnetocapacitance effect at the Curie temperature.[21] However, non-linear electrical transport studies and electroresistance in this compound are not reported so far.

We show that both the NCMO and LNMO exhibit a direct current (*dc*) induced CER effect irrespective of their distinct electronic and magnetic ground states. Our simultaneous measurements of *dc I-V* characteristics and temperature of the sample reveal that the negative differential resistance observed in these two compounds and, possibly in other manganites too, is caused by the Joule heating at high current denisties. It is also shown that, in the current regime of negligible Joule heating, the resistance can be switched between a high and a low resistance states by controlling the pulse width or period of the current excitation even if the amplitude of the current excitation is fixed. We also show multi-level resistivity switching controlled by the period and the width of pulsed current and retention behavior of memory states in these materials.

We have measured the four probe *dc* and pulsed *I-V* characteristics of bar shaped samples of the NCMO and LNMO at different temperatures using source-measure units (Keithley 2400 and Yokagawa GS610) interfaced to the Physical Property Measurement System (PPMS, Quantum Design Inc, USA). The electrical contacts were made with Ag-In alloy or Ag paste and the results were found to be identical. In order to investigate the change in temperature due to the Joule heating of the sample, a small calibrated Pt-100 resistance sensor (3mm x 3 mm x 2 mm in size) was glued to the top surface of the sample using the Apiezon-N grease for better thermal contact and its four probe resistance during the current sweep was measured using a Keithley 6221 current source and a 2182A nanovoltmeter.

Figure 1 shows the temperature dependence of the resistivity ($\rho$) of NCMO for five different *dc* current strengths ($I$ = 0.1, 1, 5, 10 and 20 mA) on the left scale, and the temperature of the sample measured by the Pt-resistance thermometer, on the right scale. The sample is non-metallic over the entire temperature for all the currents and an abrupt increase of resistivity occurs around $T$ = 250 K due to the charge-orbital ordering as reported earlier.[15] While the $\rho$ for all the values of current matches at high temperatures, it shows a dramatic decrease with increasing magnitude of the current below 150 K. The resistivity decreases as much as 80 % at 50 K as the current changes from 1 mA to 20 mA. A tendency towards the saturation with decreasing temperature and increasing current strength is clearly visible. While the Pt resistance sensor glued on the top surface of the sample and cernox sensor thermally connected to the sample holder in the PPMS recorded identical temperature for 1μA current through the sample, a significant increase in the temperature of the occurs below 150 K with increasing current as shown by the temperature readings on right scale. It can be noted that even though the base temperature, as noted by the thermometer in the PPMS, is T = 50 K when $I$ = 20 mA, the top surface of the sample is at much higher temperature ($T_{pt}$ = 117 K). This clearly indicates significant global Joule heating of the sample which was over looked in earlier works.[4,5,17] Since the sample is insulating, Joule heating can lead to the lowering of resistance in addition to non-thermal, electronic origins. For example, $\rho \approx 63$ Ω cm for 20 mA while the base temperature measured by the PPMS is 50 K, but similar magnitude of $\rho$ is obtained when $I$ = 100 μA at $T$ = 126 K. This temperature is not far from the measured temperature of the sample ($T_{pt}$ = 117 K) for $I$ = 20 mA at base temperature 50 K.

Next, we will show the current-voltage characteristics of the NCMO sample at a fixed temperature, $T$ = 100 K. The left column of fig. 2 shows (a) *dc I-V* characteristics (b) temperature of the sample and (c) resistivity for a complete current sweep (0 mA → +20 mA → -20 mA → +20 mA) in steps of 50 μA with the voltage compliance limited to 80 V. The *dc* voltage in fig. 2(a) initially increases linearly with increasing current (curve 1) but it exhibits a peak around 3 mA and then decreases for further increase in the current resulting in a negative differential resistance (NDR) regime for $I$ > 3 mA. The voltage traces a different path (curve 2, red symbols) upon decreasing the current from +20 mA, goes through a maximum around +5 mA and a minimum around -3 mA in the reverse direction and again a maximum around +3 mA in the forward direction (curve 3, blue symbols). The temperature ($T_{pt}$) of the sample (fig. 2(b)) remains nearly constant at low currents but it increases rapidly above 3 mA in the NDR regime and reaches 130 K when $I$ = 20 mA. As the current reduces, $T_{pt}$ decreases and goes through a minimum not at the origin but around -1.3 mA and increases again with increasing current in the reverse direction. The temperature of the sample also exhibits a hysteresis with forward and reverse sweeps of the current as like the *dc* voltage. The *dc* current induced electroresistance is as high as -96 % during the initial sweep from 0 mA to +20 mA as can be realized from fig. 2(c).

To confirm whether the observed NDR is caused by the Joule heating, we carried out pulsed *I-V* sweeps with different pulse widths and periods. Fig. 2(d) shows the pulsed *I-V* behavior measured with 25 ms pulse width (PW) and 1 s period (PD). The pulsed voltage sweep eliminates the NDR regime. The voltage increases quasi linearly until I = 10 mA and then shows an apparent saturation at higher currents due to the voltage compliance. As expected, the increase in the sample temperature is small (~2 K, see fig, 2(e)). The change in resistance (fig. 2(f)) is rather small below 10 mA and it decreases rapidly above 10 mA due to the apparent saturation of the voltage with increasing current.

The non linear pulsed *I-V* shows interesting evolution with increasing period for a fixed pulse width (PW = 25 ms) as depicted in fig 3(a)-(d). Each measurement was repeated twice to check for the reproducibility with about ten minutes gap between successive measurements. When the period is shortest (PD = 50 ms in fig. 3(a)), the virgin curve (0 mA → +20 mA, curve 1) lies above the curves during the down- sweep (+20 mA → -20 mA , curve 2) and the up- sweep (-20 mA → +20 mA , curve 3). When the period is increased to 0.2 s (fig. 3 (b)), the virgin curve exactly merges with the curve 3 but the hysteresis between curve 2 and curve 3 and the NDR regime persist. As the PD is increased further to 0.5 s (fig. 3(c)) and 1s (fig. 3(d)), the voltage becomes larger, the NDR regime and the hysteresis disappear. Fig. 3(e) and 3(f) show the *I-V* characteristic for longer pulse width than used in fig 3(a)-(d) and varying period. The NDR behavior re-emerges for both PW = 0.2 s and 1 s but the hysteresis decreases, the slope of the curve near the origin becomes steep and the threshold current for the NDR regimes becomes lower when the PW= 1 s and PD = 1.2 s. Such a systematic evolution of the pulsed *I-V* was not reported earlier and it suggests that the evolution of resistance between the current pulses is seriously affected not just by the pulse width but by the period as well.

The strong dependence of PW and PD on the *I-V* characteristics has motivated us to investigate the response of electrical resistivity of the NCMO to a sequence of pulses with varying PD or PW for fixed current amplitude. We have chosen, *I* = 2 mA to avoid excessive Joule heating. Fig. 4(a) shows $\rho$ on the left scale and temperature on the right scale in response to five pulse trains of 2 mA amplitude with a fixed pulse width (PW = 25 ms) but two different periods (PD = 50 ms and 0.1 s). Consecutive pulse trains consists of 200 pulses of PD = 50 ms and 200 pulses of PD = 0.1 s. After the initial drop of resistance at the beginning of the pulse train, the resistance abruptly jumps by 37 % when the PD is changed from 50 ms to 0.1 s and remains nearly unchanged until another pulse of a shorter period (PD = 50 ms) sets it to a low resistance state. The resistance can be again set to the higher magnitude by the application of the longer period pulse (PD = 0.1 s). The temperature of the sample also changes periodically from 100.9 K for PD = 0.1 s to 101. 9 K for PD = 25 ms but this 1 K change in temperature alone is insufficient to account for the resistivity change of 37 %. It is possible that the observed periodic changes in temperature with pulse sequence possibly arises from the electrocaloric effect, i.e, change in the sample temperature upon sudden, adiabatic application or removal of electrical field. The electrocaloric effect is known in some polar dielectrics and it arises from transfer of thermal energy to the lattice during polarization or absorption of thermal energy from the lattice during depolarization similar to magnetocaloric effectc.[22] Recently, it received much attention due to a temperature change of 12 K for a small bias voltage of 25 V applied to a 300 nm insulating $PbZr_{0.9}Ti_{0.05}O_3$.[23] In these experiments electric-field induced changes in the polarization is the cause of the change in the temperature of the sample. Similarly, an electric-field induced resistance change leading to the observed temperature changes is possible in our system. However, it is beyond the scope of this short communication to give a detailed discussion but we would like to stress that the observed electroresistance of 37 % or above (see below) is hard to understand by the measured temperature change of 1 K alone.

The resistive switching can also be accomplished by changing the PW for a fixed period as shown in Fig 4(b). Here, the resistivity abruptly decreases by 53 % when the pulse width is changed from 25 ms to 100 ms for a constant period (200 ms). The observed switching between two resistivity levels with varying PD or PW is not reported so far and it may offer advantage in certain cases over the conventional method of resistance switching with changing magnitude of the current. We can also obtain resistivity switching between three levels by introducing three different periods as shown in fig. 4(c). A pulse of PD = 1 s increases the $\rho$ by 44 % and a second pulse of PD = 25 ms decreases the resistance by 60 % and a third pulse of PD = 0.2 s increase the resistance by 70 % with respect to their previous states. The repetition of pulses in the same order drives the sample to the respective resistive states. Fig. 4(d) shows the effect of random switching of resistance by different PDs for a fixed PW = 25 ms. It is found that each PD drives the system into its own specific resistive state, whatsoever be the sequence. For instance, the resistivity of the sample changes from ~ 450 $\Omega$ cm to ~ 1286 $\Omega$ cm when the PD is sequentially changed from 25 ms to 3 s with intermediate resistance states attained for PD = 0.1 s, 0.2 s, 0.5 s, and 1 s. Even if the pulsed period is randomly changed after 3 s as shown in the figure 4(d), the resistive states attained exactly coincide with the previous states.

The observed current induced resistivity switching is not only confined to charge ordered manganites alone, but surprisingly can also be seen at room temperature in charge – orbital disordered LNMO. This compound also exhibits non-linear *I-V* characteristics at *T* = 300 K(not shown here) similar to the NCMO (at 100 K) but its room temperature resistivity is five orders of magnitude higher than the NCMO ($\rho$(300 K) $\approx$ 10$^{-2}$ $\Omega$ cm). The detailed temperature dependence of electrical behavior of the LNMO is beyond the scope of this communication and it will be reported elsewhere. Fig. 5(a) shows the resistance switching observed for a fixed current ($I$ = 2 mA) and period (PD = 1 s) while alternating the PW between 100 ms and 25 ms. Pulsed current of longer pulse width drives the sample into low resistive state and that with shorter pulse width transforms the low resistance state into high resistance state. The resistivity switching becomes gradual at the edges when PD = 0.2 s (fig. 5(b)) but the overall change in magnitude is rather large ($\approx$ 32 %) compared to the 9 % change when PD = 1 s (fig. 5(a)). While the PW is fixed to 25 ms (fig. 5(c)), alternating the PD between 50 ms and 100 ms results in 17 % change in resistance but the approach to saturation is rather slower than the cases 5(a) and 5 (b).

The common trends seen in the *dc*/pulsed *I-V* characteristics and pulse current induced resistance switching in the above two compounds suggest a common mechanism behind the CER other than the current induced melting of charge ordering or charge density wave excitation at low currents. We have also repeated all the above measurements with capacitance- like structure (silver electrode-oxide-silver electrode) in two probe configurations and found them qualitatively similar though they differed in magnitudes. Hence, we believe that the observed resistivity switching is of bulk effect rather than a change in the contact resistance. While the negative differential resistance possibly results from the rapid Joule heating of the sample at high current densities, the CER effect observed by variations in the pulse period or the pulse width can not be attributed to thermal effects alone. We, therefore, suggest the following scenario. It is known that electrical transport in these materials is dominated by thermally assisted hopping of self-trapped electrons (polarons) in their own potential well due to the strong interactions with surrounding lattice and dynamical Jahn-Teller interaction. The trapped electron stays at the bottom of the potential well unless it gains sufficient energy to overcome the barrier. A wider pulse width or shorter interval between pulses supply sufficient power and can de-trap the electron from the potential well leading to a lower resistance. Whereas, if the pulse width is narrow or the time interval between successive pulses is sufficiently long, only fewer electrons are detrapped and it increases the resistance as clearly demonstrated in the pulsed current sweep (fig. 4 (a)-(d)). Another possibility is that pulsed current causes non equilibrium current density which relaxes slowly towards equilibrium that depends on the pulse width and period. A systematic study in the regime of wide range of pulse width, particularly in sub millisecond pulse range can further shed light on the mechanism of the observed effect.

In summary, our investigations of direct and pulsed current–voltage (*I-V*) characteristics in charge-orbital ordered $Nd_{0.5}Ca_{0.5}MnO_3$ and charge disordered $LaNi_{0.5}Mn_{0.5}O_3$ reveal that negative differential resistance observed at higher currents in these materials are caused by significant Joule heating. It was found that pulsed current sweep with shorter pulse widths or longer periods result in higher resistance than the *dc* current sweep and disappearance of the negative differential resistance regime. It is shown that resistance of these materials can be switched between high and low resistive states by controlling the pulse widths and periods even if the amplitude of the pulsed current is fixed. We attribute the observed pulse width and period dependent electroresistance to the trapping and de- trapping of charges within the bulk of the material. The memory retention of states driven by the pulse period and pulse width is also very effective in our sample, which is independent of the sequence of applied pulses. These key features may make them suitable for applications in future RRAM.

R. M acknowledges MOE and NUS for support of this work through grants MOE/NUS/AcRF-Tier1- R-144-00-167-112.

**Figure captions:**

Fig. 1 Temperature dependence of the resistivity of $Nd_{0.5}Ca_{0.5}MnO_3$ for different values of the *dc* currents. The *x*-axis shows the base temperature of the sample measured by the cernox sensor installed in the cryostat close to the sample holder. The *y*-axis on the right scale shows the temperature measured by the Pt-sensor glued on the top of the sample. Note the disagreement between these two temperatures for higher current strengths below 150 K due to the Joule heating of the sample.

Fig. 2 Current (*I*) dependence of (a) the *dc* voltage (*V*) of $Nd_{0.5}Ca_{0.5}MnO_3$ at 100 K during the *dc* current sweep $I = 0 \rightarrow +20$ mA (black), $+20$ mA $\rightarrow -20$ mA (red) and $-20$ mA$\rightarrow +20$ mA (blue), (b) the sample temperature measured by the Pt-resistance and (c) resistivity. The right column shows the respective quantities during the pulsed current sweep. Note that the induced voltage is large and the increase in temperature is small in the pulsed mode than the *dc* mode.

Fig. 3 (a)-(d) Evolution of the non-linear pulsed *I-V* characteristics of $Nd_{0.5}Ca_{0.5}MnO_3$ at 100 K with fixed pulse width (PW = 25 ms) but increasing periods (PD). The values of PDs and PW are mentioned in the graph. The virgin curve ($0 \rightarrow +20$ mA, black symbol) is clearly separated from other down- and up- sweep curves in fig. 3(a). 3(e) and 3(f) show the *I-V* behavior with much longer pulse width (PW > 25 ms) and periods.

Fig. 4 (a) Resistivity switching (red symbols) at a constant current ($I = 2$ mA) as a function of pulse numbers for $Nd_{0.5}Ca_{0.5}MnO_3$ induced by variations in (a) pulse periods for a fixed pulse width (PW = 25 ms) and (b) pulse widths for a fixed PD = 0.2 s. Arrows indicate the start of a new pulse train and the numbers indicate PD or PW. The periodic changes in temperature of the sample is shown on the right scale (blue symbols). (c) Tri-level resistivity switching for three different periods for PW = 25 ms and (d) random switching of resistance induced by sequence of pulses of varying period. It is to be noted that each PD or PW drives the system in to its own specific stable resistive state irrespective of the order. Arrows indicate the start of a train with new PD or PW.

Fig. 5　Resistivity switching in $La_2NiMnO_6$ at T = 300 K induced by variations in pulse widths for (a) PD = 1 s, (b) PD = 0.2 s and for varying periods for (c) PW = 25 ms.

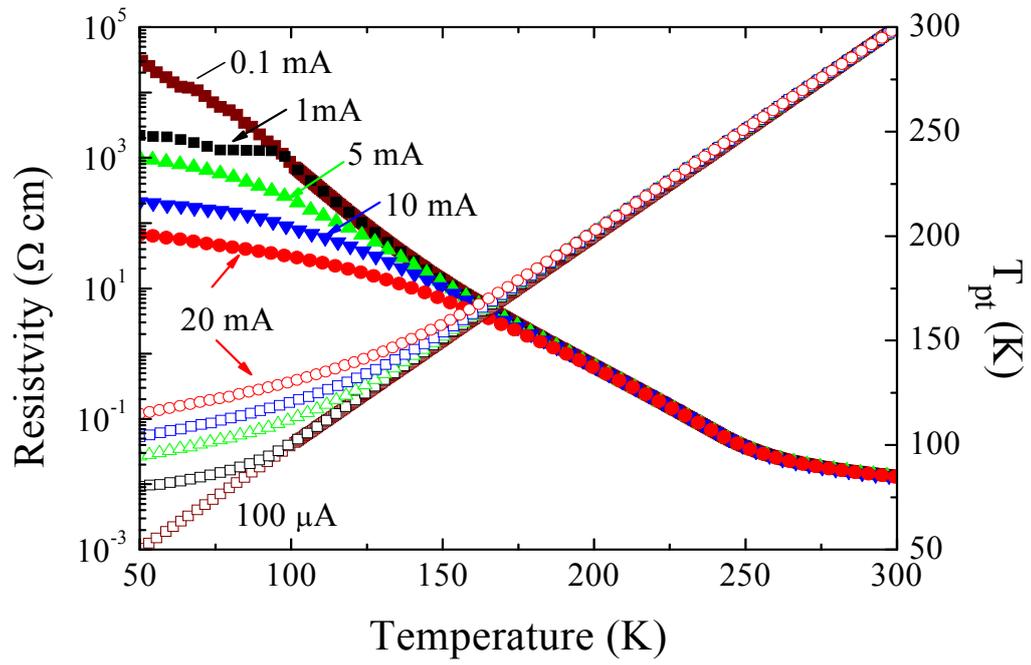

Fig. 1

A. Rebello et al.

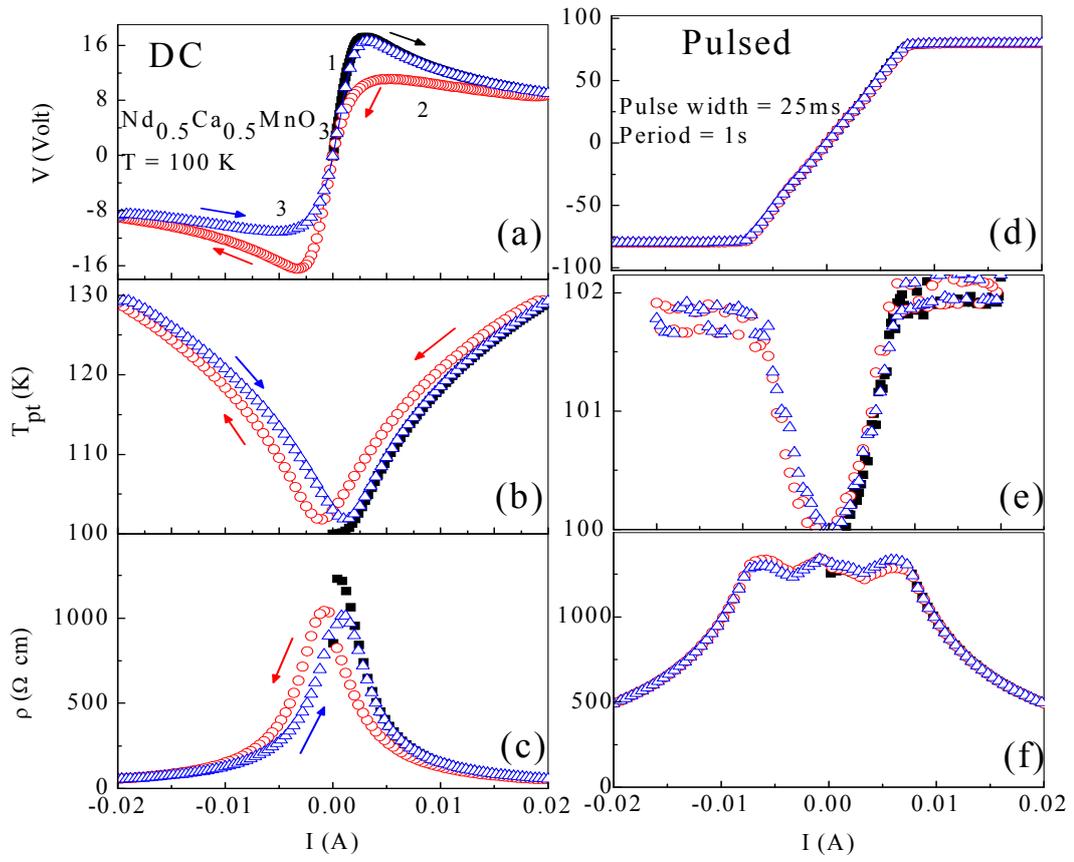

Fig. 2
A. Rebello et al.

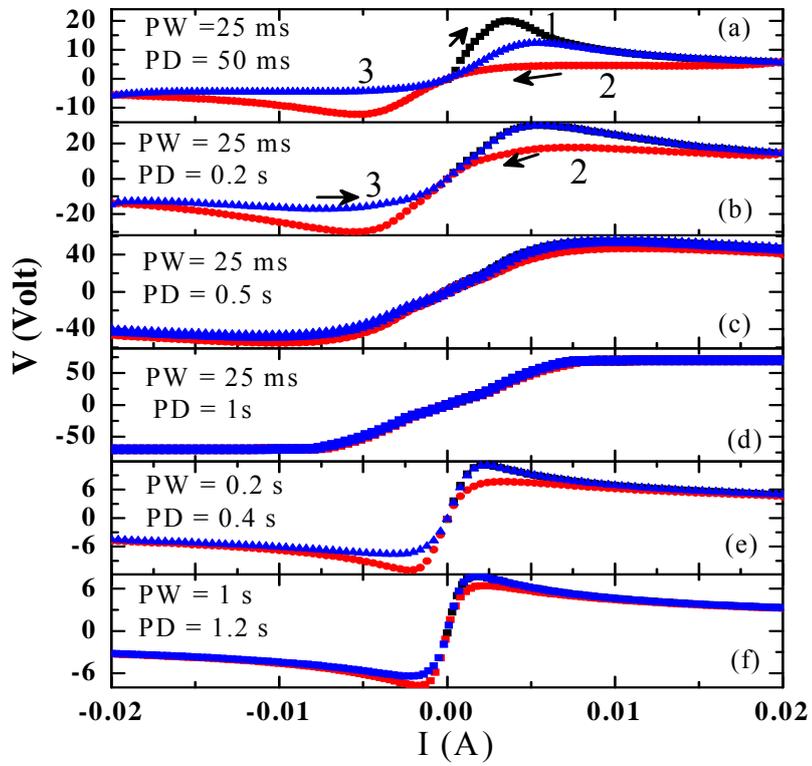

Fig. 3

A. Rebello et al.

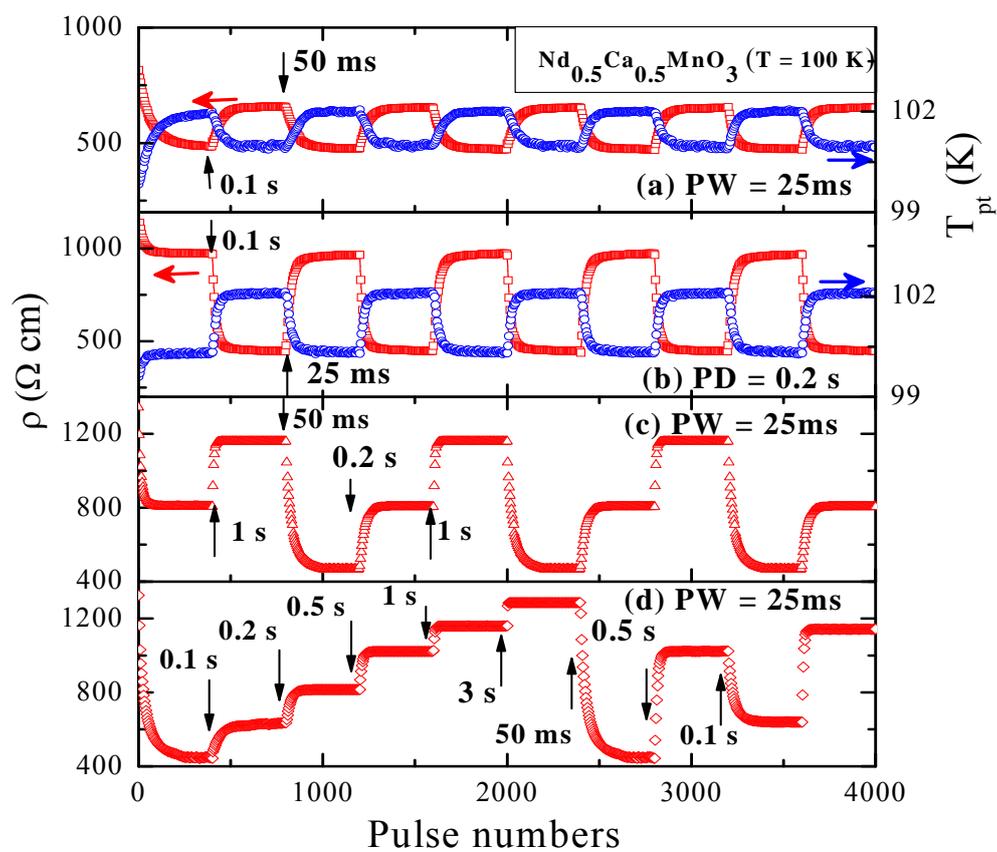

Fig. 4

A. Rebello et al.

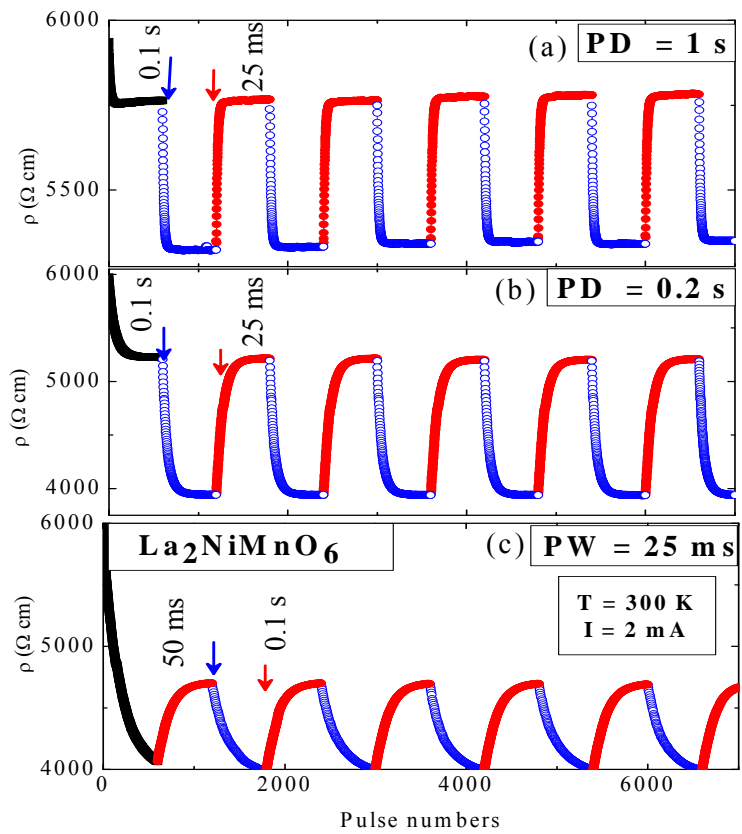

Fig. 5

A. Rebello et al.